\documentclass[natbib]{svjour3_gks}                     
\smartqed  
\usepackage{graphicx}
\usepackage{color}

\newcommand{\degr} {$^\circ$}
\newcommand{\ttwopi}{$t_{2\pi}$}
\newcommand{\muas}{$\mu''$}
\newcommand{\micron}{$\mu$m}

\newcommand{\skinnerpfla}{2001A&A...375..691S}
\newcommand{\casha}{2000Natur.407..160C}
\newcommand{\cashb}{2003ExA....16...91C}
\newcommand{\willingale}{2004SPIE.5488..581W}
\newcommand{\shipley}{2003SPIE.4851..568S}
\newcommand{\cashc}{2003SPIE.4852..196C}
\newcommand{\clauser}{1992ApPhB..54..380C}
\newcommand{\gerlich}{2007NatPh...3..711G}
\newcommand{\momose}{2004AIPC..716..156M}
\newcommand{\pfeiffer}{2008NatMa...7..134P}
\newcommand{\skinnerao}{2004ApOpt..43.4845S}
\newcommand{\broderick}{2009ApJ...697.1164B}
\newcommand{\jovanovi}{2009arXiv0903.0978J}
\newcommand{\wuwang}{2007MNRAS.378..841W}
\newcommand{\iwasawa}{2004MNRAS.355.1073I}
\newcommand{\vaughan}{2008MNRAS.390..421V}
 \newcommand{\white}{2000Natur.407..146W}

 \journalname{To be published in ``Experimental Astronomy"   }
 \doi{DOI: 10.1007/s10686-009-9175-4 }
 \textmessage{The original publication will be available at www.springerlink.com.}
\begin{document}

\title{X-ray interferometry with transmissive beam combiners for ultra-high angular resolution astronomy
}


\author{G.~K.~Skinner         \and J. ~F.~Krizmanic
}


\institute{G.~Skinner \at
              NASA-GSFC \& CRESST,  Greenbelt, Md 20771, USA, \& Dept. of Astronomy, Univ, Md, College Park, Md 20742, USA \\
              Tel.: +301-286-1350\\
              \email{skinner@milkyway.gsfc.nasa.gov}          \\
          J. ~Krizmanic \at
              NASA-GSFC \& CRESST,  Greenbelt, Md 20771, USA, \&  Universities Space Research Association 
}


\maketitle

\begin{abstract}
Interferometry provides one of the possible routes to ultra-high angular resolution for X-ray and gamma-ray astronomy. Sub-micro-arc-second  angular resolution, necessary to achieve objectives such as imaging the regions around the event horizon of  a super-massive black hole at the center of an active galaxy, can be achieved if beams from parts of the incoming wavefront separated by 100s of meters can be stably and accurately brought together at small angles. One way of achieving this is by using grazing incidence mirrors. We here investigate an alternative approach in which the beams are recombined by optical elements working in transmission. It is shown that the use of diffractive elements is a particularly attractive option. We report experimental results from a simple 2-beam interferometer using a low-cost commercially available profiled film as the diffractive elements. A  rotationally symmetric  filled (or mostly filled)  aperture variant of such an interferometer, equivalent to an X-ray axicon, is  shown to offer a much wider bandpass than either a Phase Fresnel Lens (PFL) or a PFL with a refractive lens in an achromatic pair. Simulations of an example system are presented.
   
\keywords{First keyword \and Second keyword \and More}
 \PACS{ 07.85.Fv \and 07.85.-m \and  41.50.+h   \and 87.59.-e \and 95.55.Ka}
\end{abstract}

\section{Introduction}
\label{sect_intro}

To image the space-time around the event horizon of  super-massive black holes (SMBHs)  such as those believed to be at the core of many galaxies requires an angular resolution of the order of 0.1--1 micro arc seconds (\muas), even in the most favorable cases \citep{\white}. Even with wavelengths corresponding to the X-ray and gamma-ray bands,  an imaging system must have a large aperture to achieve resolution of this order. At very  high photon energies ({e.g.} greater than 100 keV) filled aperture lenses of the required  diameter  may be just practical \citep{\skinnerpfla}, but at lower energies much larger apertures would be needed.  

VLBI radio observations have been suggested as a possible way of achieving this objective \citep[e.g.][]{\broderick, \jovanovi} but even with apertures of the size of the earth, and at sub-mm wavelengths, they are limited to angular resolutions $\sim$10 \muas, and hence to studying our own galactic nucleus  and perhaps those of a few nearby galaxies.  

It has been argued \citep{massim.proposal} that ultra-high angular resolution observations in the Fe  K-alpha emission line (rest energy 6.7--6.9 keV for He-like or H-like Fe)  are particularly relevant. 
This line has been reported with what appear to be redshifts, broadening, and line-profile suggesting that it can originate very close to a SMBH\footnote {Note : other explanations of the observations are possible, \citep[e.g.][]{\vaughan}.}.  There are even indications that periodically varying redshifts corresponding to flares orbiting close to a SMBH have been seen \citep{\iwasawa}. Simulations exist of just how SMBHs should appear at these energies if and when imaging with adequate angular resolution becomes possible \citep[e.g.][]{\wuwang}.
 
However  at, for example,   6--7 keV,  the Rayleigh limit implies   a diameter of $d\sim$500~m for an angular resolution of 0.1\muas . Filled apertures of this size are neither practicable nor necessary to collect adequate signal. Thus interferometric approaches are likely to be preferred and a space mission, MAXIM, with this capability has been studied \citep{\cashc}. 

We discuss in Sections \ref{sect_designs}--\ref{sect_multibeam} how diffractive beam combiners might form a key part of a mission with these objectives. 
Even where the photon energy and the desired angular resolution imply a size for which a filled aperture is conceivable, treating a diffractive optic as an interferometer can lead to a design that has a far wider bandpass than a  Fresnel lens, even one that is made `achromatic' by combination with a refractive  component. In Section \ref{sect_symmetric}   we show how this approach  leads to a design that is effectively an X-ray axicon and present simulations comparing the performance of an example X-ray axicon design with that of lenses.

\section{Designing an X-ray interferometer}
\label{sect_designs}

\citet{\cashb} has reviewed many of the issues that arise in trying to design an X-ray interferometer for astronomical applications.
Basically, an interferometer  has to be able to combine the beams from multiple sub-apertures that are spread across the wavefront by distances up to $d$, the longest baseline that is required. The distance $s$  between fringes must not be too fine to measure,  setting a limit on  the angle $\psi=\lambda/s=hc/Es$  at which the beams converge ($\lambda$ being the wavelength, $E$ the photon energy). \citet{\casha} have described a system  in which beams are combined using `periscopes' containing four grazing incidence mirrors and have demonstrated the formation of fringes in a simple set-up. \citet{\willingale} has discussed a variant of this arrangement in which a slatted mirror is proposed as a `semitransparent' X-ray mirror to allow the beams to be combined at small $\theta$ in a compact arrangement.

\begin{figure}[htbp]
\begin{center}
\includegraphics[width=82mm, trim= 100 0 50 0 ]{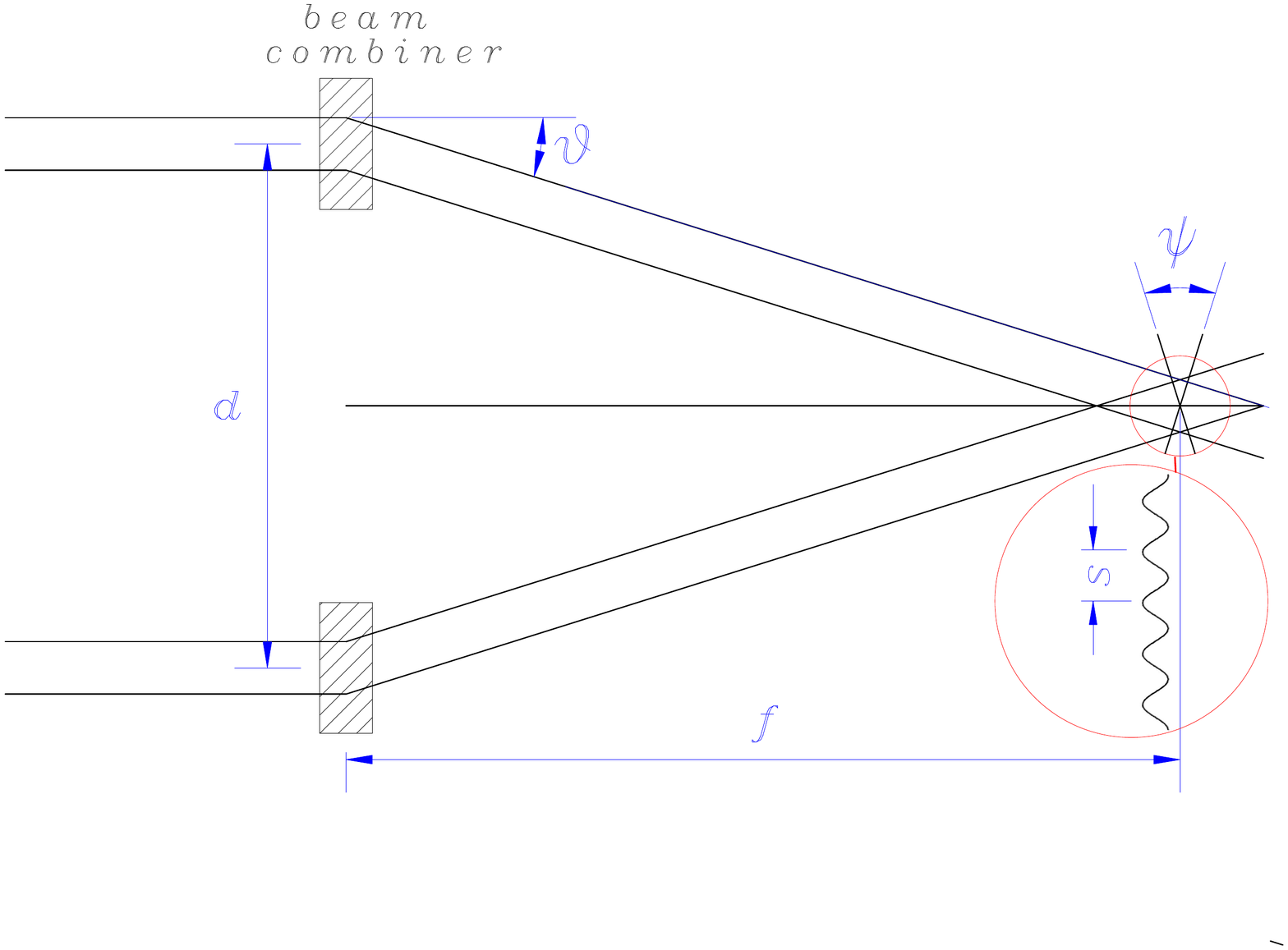}
\vspace{-12mm}
\caption{A generic simple 2-beam interferometer.}
\label{fig_generic}
\end{center}
\end{figure}

\begin{figure}[htbp]
\begin{center}
\includegraphics[width=85mm, trim= 0 200 0 50 ]{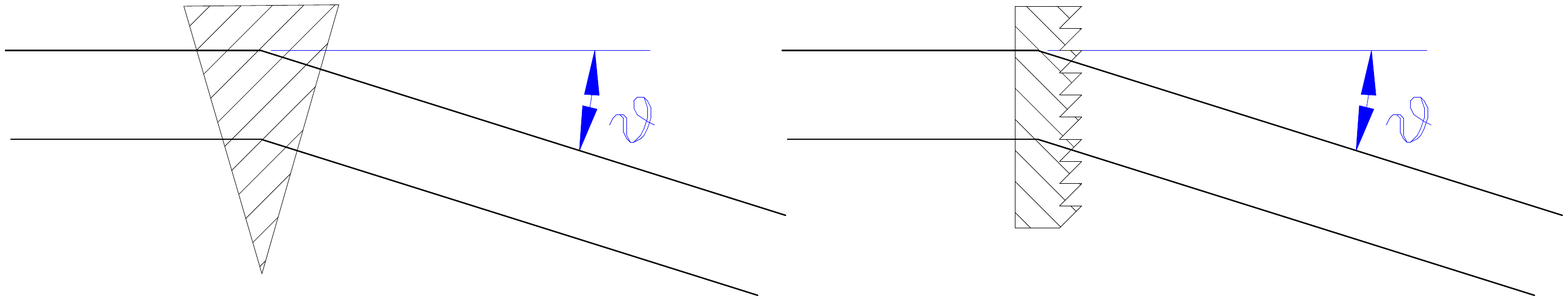}
\caption{(a) Use of a Prism as a refractive beam combiner.  Note the direction of the deflection, resulting from a refractive index less than unity. Two such combiners are equivalent of a Fresnel bi-prism. (b) Use of a  blazed phase grating as a beam combiner.}
\vspace{-15mm}
(a)\hspace{37mm}(b)\hspace{20mm}
\vspace{15mm}
\label{fig_refr_diffr}
\end{center}
\end{figure}

 We consider here an interferometer that is as conceptually simple as possible (Fig.~\ref{fig_generic}). Two beam combiners divert the incoming radiation to arrive on the same detector. For a sufficiently large $f/d$ the beams converge at a small angle $\psi$  and the angle $\theta=\psi/2$ by which each of the beams has to be diverted is also small. Although a two-beam interferometer is shown, more beams could be combined by having more `beam-combiners', as discussed in \S \ref{sect_multibeam} below.  One way of regarding the generalized configuration shown in Fig. \ref{fig_generic} is that two virtual images of every point in the object illuminate the detector and produce  fringes in a  form of Young's slits configuration. 

The beam-combiners could be simple grazing incidence mirrors but in this case the orientation of each  would have to be extremely stable. With the four mirror `periscopes'  mentioned above this problem is very much reduced and to a good approximation only internal stability within a periscope unit is required.    
With mirror beam combiners, the region in which the fringes are formed is dictated by the geometry and, except for broadening by diffraction, is independent of energy. The fringe spacing $s$ is proportional to $\lambda$.

 Refractive indices at X-ray energies are not quite equal to unity, so  the beam-combiners could use refraction. Two prism  combiners   (Fig. \ref{fig_refr_diffr}a) then form the equivalent of a Fresnel bi-prism. X-ray refractive indices are usually specified by $\delta (=1 - \mu)$, which is a small positive number.
For a prism angle $\alpha$ and for small $\delta$   the fringe spacing is simply given by
\begin{equation}
 s\sim \frac{\lambda}{4\: Tan(\alpha/2)\delta} \sim \frac{\lambda}{2\alpha\delta},
 \label{eqn_refr_spacing}
 \end{equation}
 the second approximation applying if $\alpha$ is also small.
 In a given material, for photon  energies well away from absorption edges,  $\delta \propto  \lambda^{2}$. Thus $s$  varies as $\lambda^{-1}$. 
The arrangement is not very practicable because  in practice $\delta$ is extremely small and the prisms have to be very thick and so have low (and progressively varying) transmission.

Consider now what happens if the beam combiners are neither mirrors nor prisms but diffraction gratings. Fig. \ref{fig_refr_diffr}b  illustrates the use of  blazed phase gratings. If the grating pitch is $p$, then for small angles, $\theta=\lambda/p$,  so
\begin{equation}
s=\lambda/\psi=\lambda/2\theta=p/2.
 \label{eqn_diffr_spacing}
 \end{equation}
Thus in this case the fringes are {\it achromatic}. Furthermore, at many X-ray energies the absorption within the diffractors can be negligible and if the blazed profile is close to ideal  then almost all of the incident energy goes into the two interfering beams.

\section{Relationship to Talbot-Lau interferometry}
\label{sect_talbot}

The formation of fringes with diffractive beam combiners is closely related to the Talbot effect, which is usually described the formation of a  self-image behind  a periodic structure such as a diffraction grating. The Talbot effect, and the closely related Lau effect in which a second grating is placed in the region where the self-image would be formed, have been widely used recently as the basis of systems for phase-contrast imaging with X-rays \citep[{e.g.}][]{\momose, \pfeiffer}. The main  difference between the interference pattern used here and that produced by the classical Talbot effect   is that true self-imaging of a grating by the Talbot effect only occurs at certain wavelength dependent distances. This is because higher diffraction orders must add to the complex amplitude with the correct phase in order to form the self-image of an object whose form contains harmonics of the basic period.  The {\it periodicity} of the fringe structure behind a single grating  is always that of the grating, independent of wavelength and distance,  but its {\it form} is wavelength dependent  and only corresponds to that  of the grating at multiples of the Talbot distance $p^2/\lambda$. We have been considering here two perfectly blazed subsections instead of a single continuous grating with arbitrary structure.  So only diffraction orders  $+1$ and $-1$ are involved and the fringes are sinusoidal, achromatic, and found at all distances where the beams cross.

Whereas in the classical Talbot effect it is the transmission profile of the grating that is imaged, 
in common with much work using in Talbot interferometers for X-ray, atomic waves or even molecular beams \citep[{e.g.}][]{\clauser, \gerlich}, the gratings used here are {\it phase-}gratings but  are imaged as {\it amplitude} fringes.

\section{ Effect of misalignment}
\label{sect_misalignment}

One reason that the more complex `periscope' configurations of mirrors mentioned in \S \ref{sect_designs}  are an improvement over the use of single mirrors is that they can be relatively immune to the changes in the orientation of periscope assembly \citep{\shipley}.  Diffractive  beam combiners, being thin and used in transmission, are  intrinsically remarkably robust in this sense.
\begin{figure}[htbp]
\begin{center}
\includegraphics[width=65mm, trim= 0 0 0 0  ]{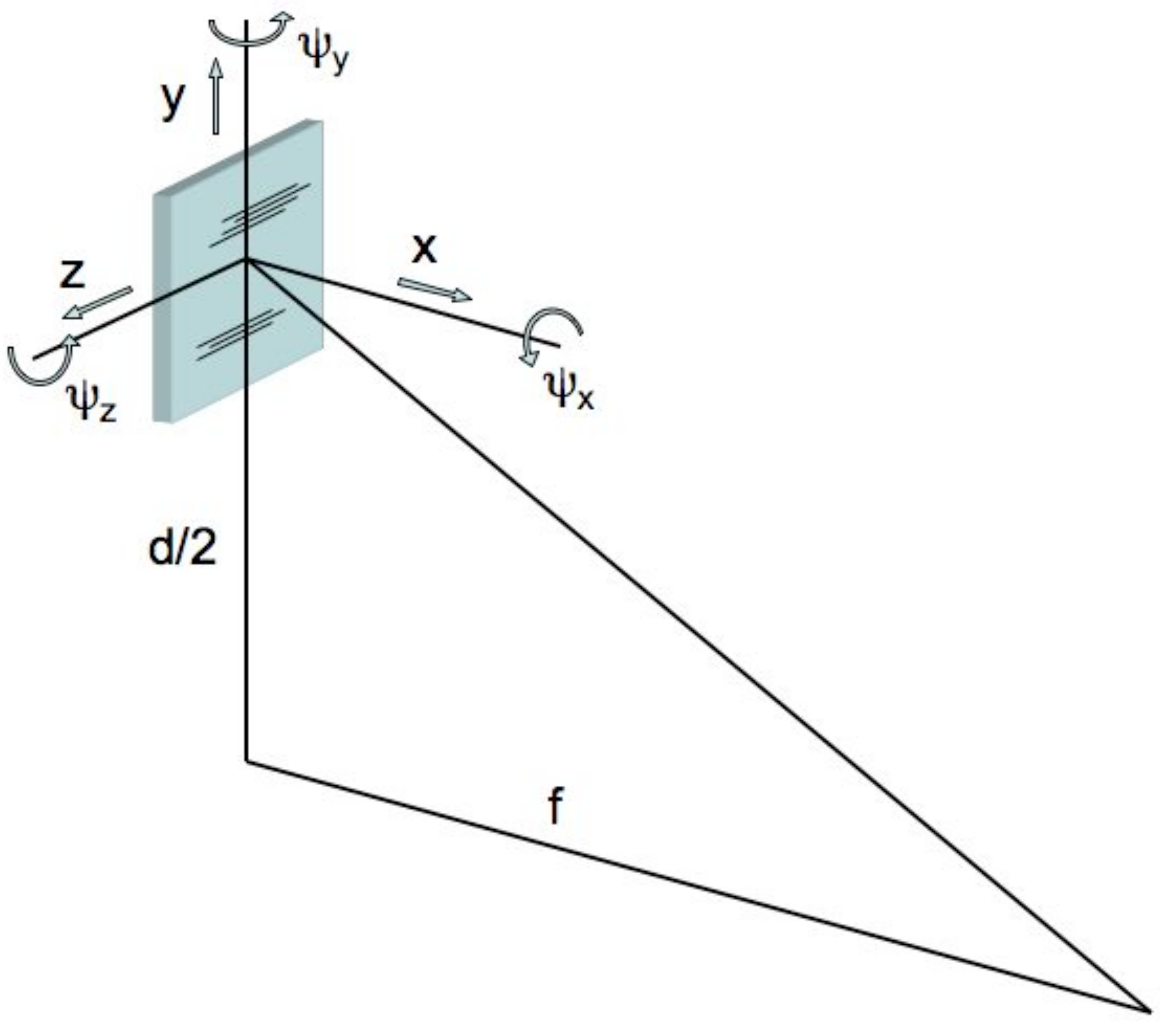}
\vspace{-2mm}
\caption{Definition of axes for consideration of alignment tolerances.}
\label{fig_axes}
\end{center}
\end{figure}
Of the six coordinates defined in Fig. \ref{fig_axes}, only a change in $y$ has a major effect on the fringe system. Tolerances corresponding to $\lambda/10$ precision are listed in Table \ref{table_tol}, along with values for an example system. All are readily achievable with existing technologies, with the exception of that on the off-axis distance $y$.\footnote{In fact with a 2 beam interferometer not even the value of $y$ is critical if only the fringe amplitude (not phase) is required.  $y$ must simply be  stable.} 

 \begin{table}[htdp]
\caption{Approximate tolerances corresponding to errors of $\frac{1}{10}$  fringe  for an interferometer with  diffractive (or refractive) beam combiners. Example values are calculated for a system  assuming:  Wavelength $\lambda= 0.185$ nm (photon energy 6.7 keV), diffractor size $\Delta y \times \Delta z$~=~1~m$^2$, diffractor separation $d=$370 m, diffractor pitch $p=$50 \micron, detector size $\Delta d$=25 cm, focal length $f=50000$ km.  The material is assumed to be Kapton, for which $t_{2\pi}$=27 \micron. The fringe spacing $s=$25 \micron\ corresponds to an angular resolution of 0.1 \muas. For definitions of axes see Fig. \ref{fig_axes}.  The tolerance on $y$, shown in bold, is the only one that presents a serious technical challenge.}
\label{table_tol}
\begin{center}
\begin{tabular}{|c|c|c|c|}
\hline
Parameter & Tolerance & Example         &  Units \\
                     &                   &   value              &            \\
\hline
x                                 &   $f\: \Delta y /d $                                                                                                & $>$100            &   km      \\
{\bf y}                          &   {\bf{p/20}}                                                                                                          &       {\bf 2.5}       & {\bf \micron }       \\
z                                 &   $(\Delta z - \Delta d)/2$                                                                                   &       0.4           &   m                                    \\
$\psi_x$                   &   $p / (10 \Delta d )$                                                                                           &      4                 &    arc seconds                     \\
$\psi_y $                  &  $Sin^{-1}\left( \sqrt{1-0.9^2}\right)$                                                                &     11          & degrees                        \\
$\psi_z$                   &     $\sqrt{\frac{p}{5 d}} $                                                                                       &       35           &   arc seconds                   \\
Diffractor profile      &                                $t_{2\pi}/10 $                                                                        &           2.7                &         \micron                             \\
\hline
\end{tabular}
\end{center}
\label{table_tolerances}
\end{table}%

\section{Laboratory demonstrations}
\label{sect_results}

\begin{figure}[htbp]
\begin{center}
\includegraphics[width=65mm, trim=200 200 200 150 ]{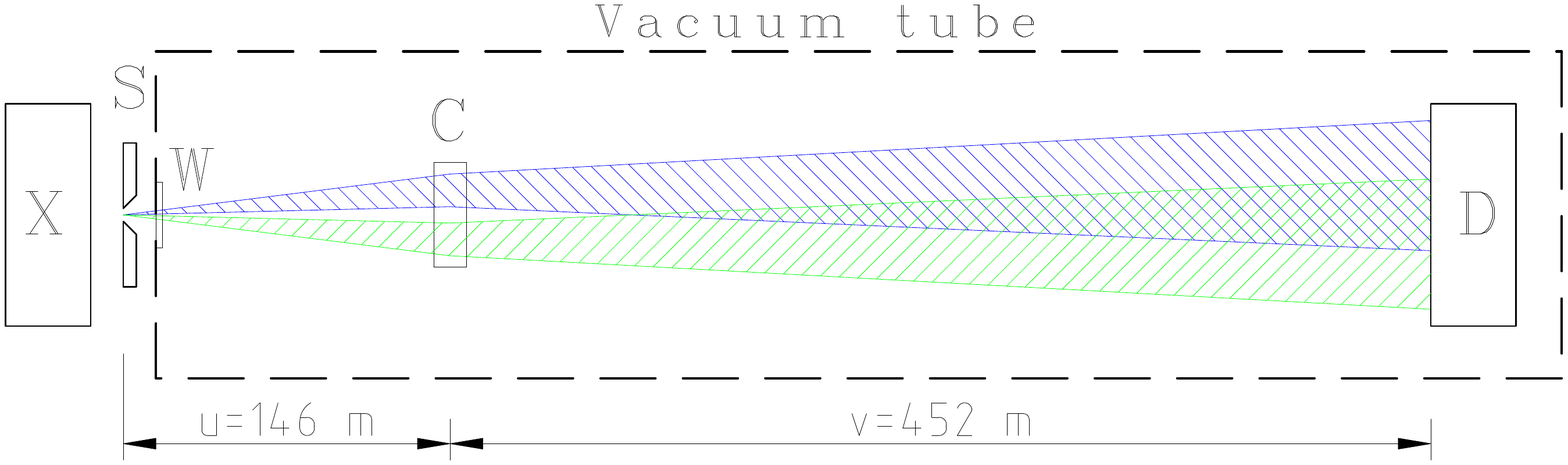}
\vspace{3mm}
\caption{The setup used for the tests at the GSFC X-ray interferometry testbed.  X~: Oxford instruments X-ray tube. S~: 5 \micron ~ Tungsten slit. W~: Beryllium window. C~: beam combiners. D~: Roper X-ray camera. The drawing is diagrammatic only -- the vacuum tube is 25 cm diameter over most of its length but with optics chambers up to 1.5 m in diameter.}
\label{fig_test_setup}
\end{center}
\end{figure}

We have performed some simple laboratory demonstrations of the principles of the proposed techniques using the  NASA-GSFC X-ray interferometry testbed\footnote{\textcolor{blue}{http://lheawww.gsfc.nasa.gov/\~{}kcg/beamline/home.html}}. 
The configuration is shown in Fig.~\ref{fig_test_setup}.
Beam combiners analogous to those in Fig.~\ref{fig_refr_diffr} were placed at distance $u=$146 m  from the source end of the 0.6 km vacuum tube. The  fringes were detected with a Roper PI-LCX/LN camera with a liquid nitrogen cooled  EEV CCD-47-10 detector having 1024$\times$1024 13 \micron\ square pixels, $v=452$m away at the far end. The source was an Oxford Instruments Apogee microfocus X-ray generator with a 5~\micron ~ Tungsten slit  between the tube window and the entrance widow of the testbed vacuum tube.

\begin{figure}[htbp]
\begin{center}
\includegraphics[width=90mm, trim=0 0 0 0 ]{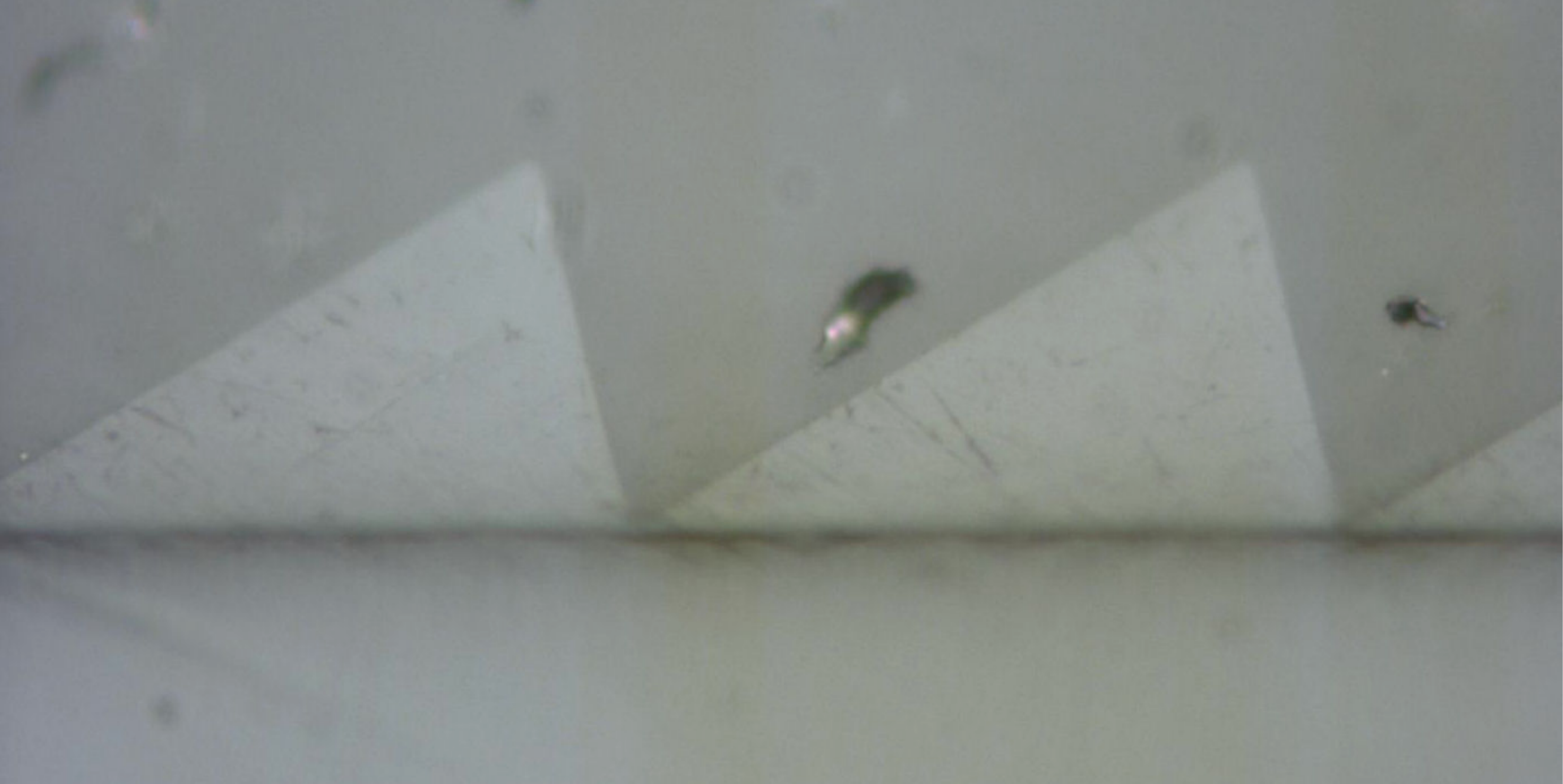}
\caption{Photomicrograph of a cross-section through the Vikuiti IDF film. The particles above the film are inclusions in the encapsulant used in preparing the section.   }
\label{fig_vikuiti}
\end{center}
\end{figure}

\begin{figure}[htbp]
\begin{center}
\includegraphics[width=100mm, trim=80 0 0 0 ]{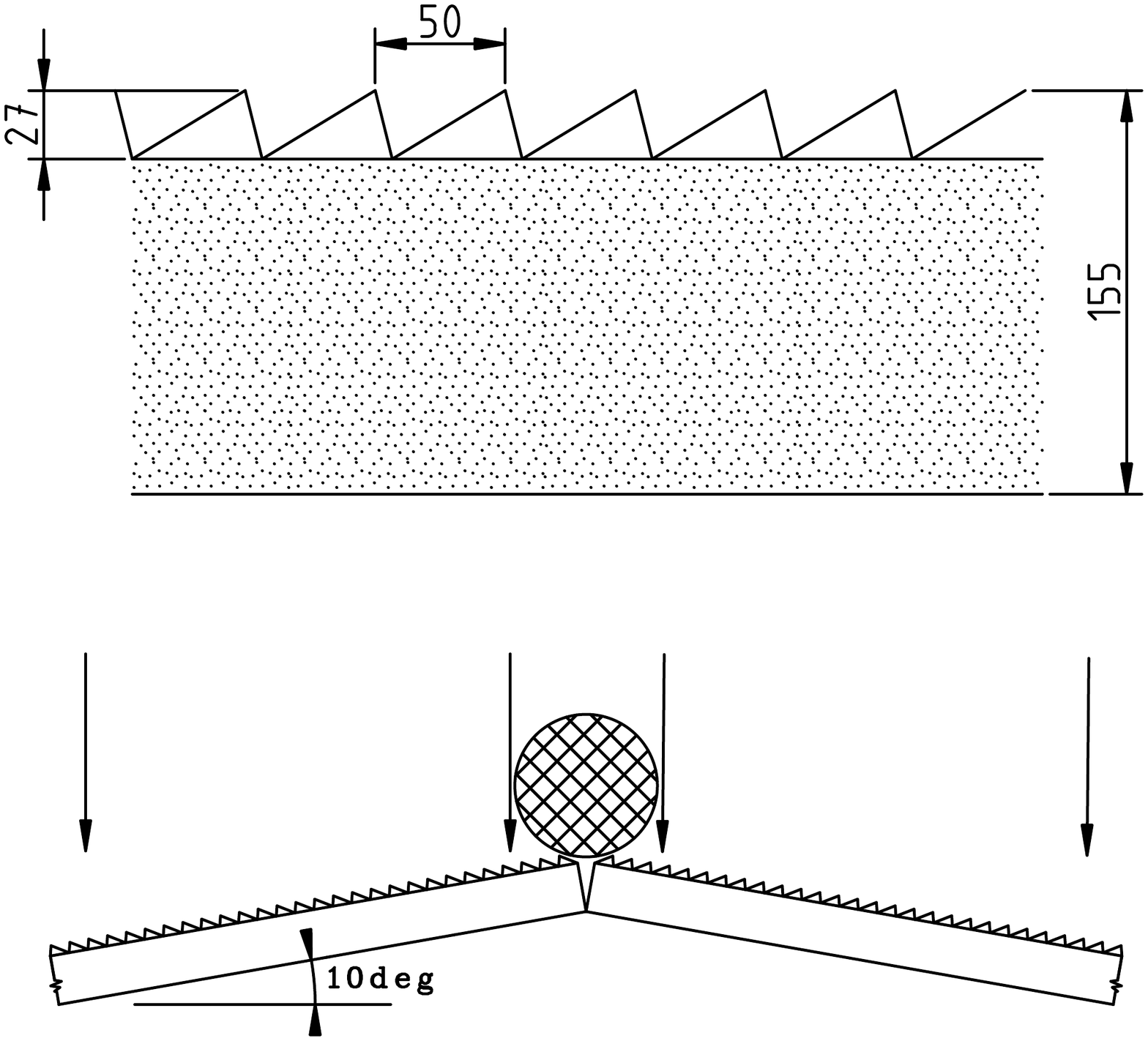}
\caption{ (a) Dimensions of the Vikuiti film (in \micron). (b) Configuration of the two beam combiners. Each piece of film was tilted by 10\degr\ to improve the mark-space ratio of the profile as seen from the beam direction. A 380 \micron\ diameter wire blocked the direct beam through the poorly defined central region.}
\label{fig_vikuiti_config}
\end{center}
\end{figure}

Plastic film with a profile very close to ideal for use as a diffractive beam combiner is produced commercially for quite different purposes. 3M Vikuiti IDF film is intended to improve the visibility of displays such as aircraft instruments that will be viewed from an angle. It consists of a Polyester substrate with a Modified Acrylic Resin surface layer with a profile (Figs. \ref{fig_vikuiti}, \ref{fig_vikuiti_config}a) that is almost ideal for use as a blazed diffraction grating in the 5--10 keV X-ray band. The `mark-space' ratio of the profile was improved by tilting the films through $\pm$10\degr\  as shown in Fig. \ref{fig_vikuiti_config}b.

\begin{figure}[htbp]
\begin{center}
\includegraphics[width=90mm, trim= 0 0 0 0 ]{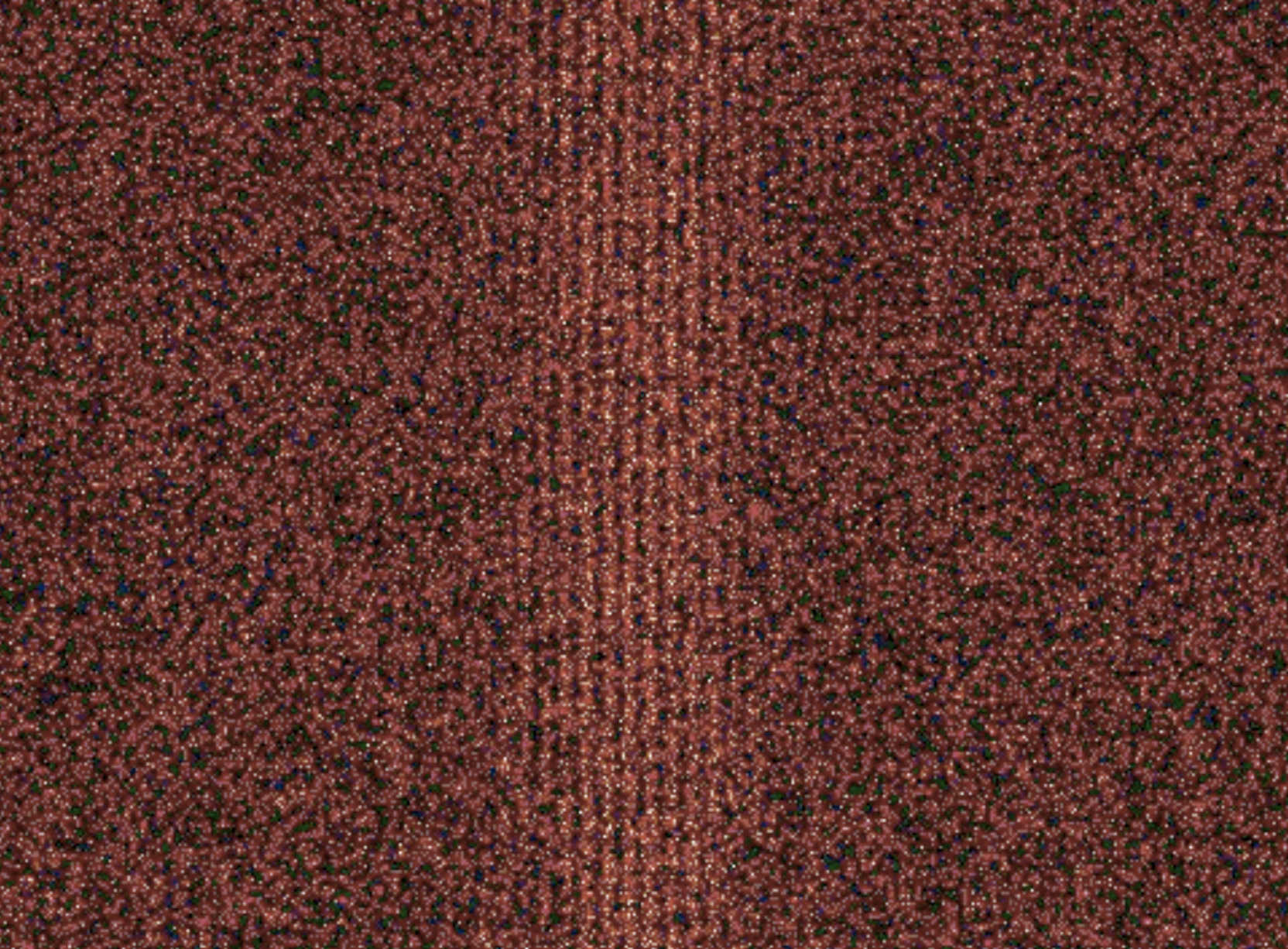}
\caption{Fringes obtained with diffractive beam combiners as described in the text. The events recorded are  a mixture of Cu K$\alpha$ (8.04 keV),  Copper K$\beta$ (8.90 keV)  and continuum radiation. }
\label{fig_fringes_image}
\end{center}
\end{figure}

\begin{figure}[htbp]
\begin{center}
\includegraphics[width=90mm, trim= 0 0 0 0 ]{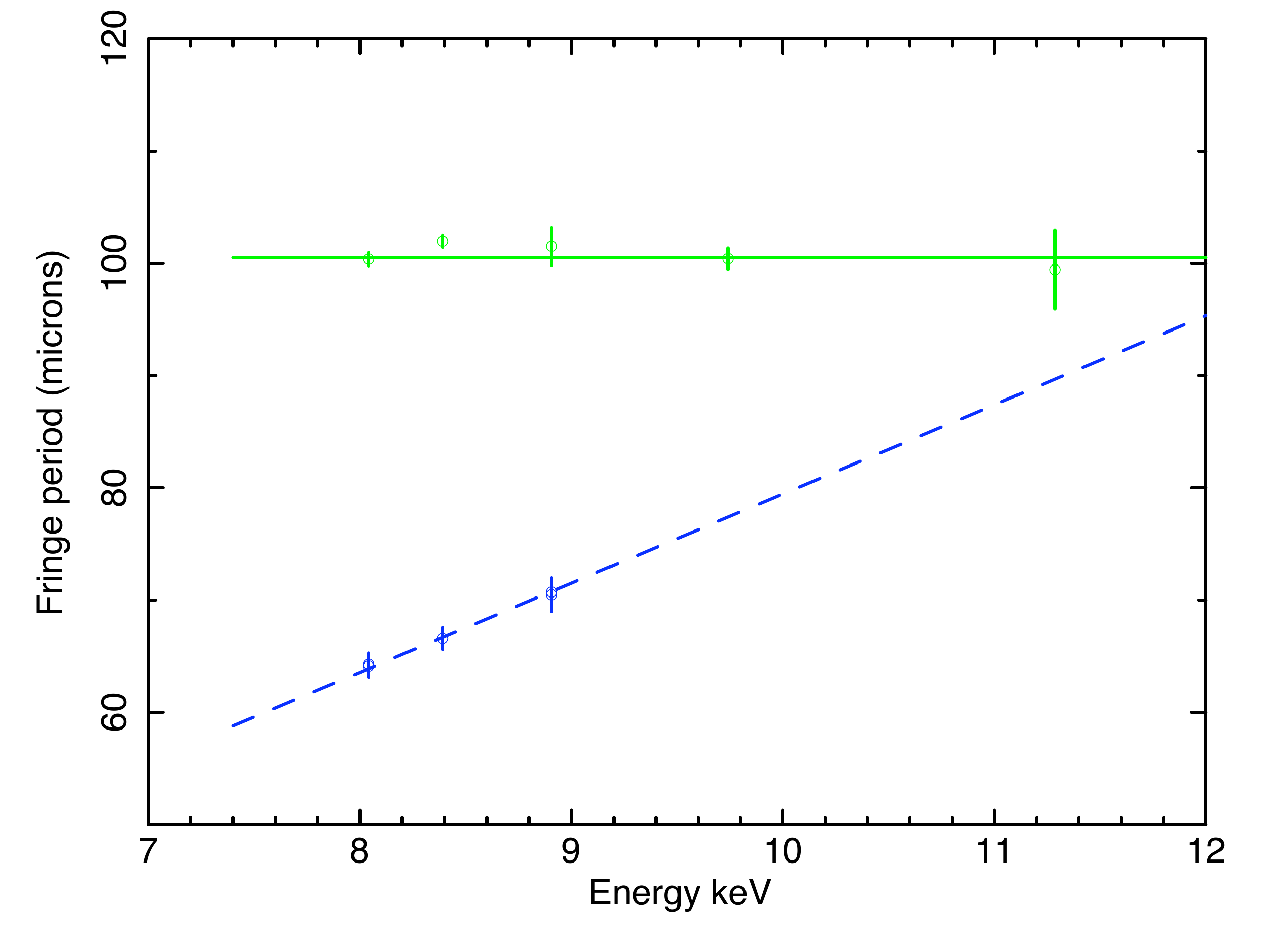}
\caption{ Measured fringe period as a function of energy. Green: with diffractive beam combiners. Blue: with refractive (prism) beam combiners for comparison. The lines show the results expected from the equations in this paper.}
\label{fig_period_v_energy}
\end{center}
\end{figure}

Fig. \ref{fig_fringes_image} shows fringes obtained using the beam combiner configuration shown of Fig. \ref{fig_vikuiti_config} with a Copper target X-ray source. During the overnight data collection there was  drift in the system alignment corresponding to about 40 \micron\  at the source. A correction was applied based on  spline fit to the shifts in the fringe pattern seen in $\sim$hour  long subsets of the data. The finite distance  of the source leads to a magnification of the fringes by a factor $(u+v)/u=4.1$  compared with Eqn. \ref{eqn_refr_spacing}. Allowing for the projection effect of the tilted diffractors, this leads to an expected periodicity of 101.2 \micron, in close agreement with the measurements. Similar results were obtained with a Tungsten target tube from which most of the radiation was in the L shell fluorescence lines (8.3--11.3 keV). By selecting events based on pulse height from the detector (energy resolution 280 eV), radiation from different lines or close groups of lines can be separated. Fig. \ref{fig_period_v_energy}  shows that the fringes are indeed achromatic.

For comparison we also obtained fringes using two  60\degr\ lucite  prisms, touching at their apexes, as beam combiners.  As was noted in \S \ref{sect_designs} the fringes are in this case expected to be chromatic, with $s\propto\lambda^{-1}$. The blue line in Fig. \ref{fig_period_v_energy} shows a prediction using values of  $\delta$ calculated using the data based on the work of Henke\footnote{\textcolor{blue}{http://henke.lbl.gov/optical\_constants}}  and assuming that the prisms consist of  pure $C_5H_8O_2$  with a density of 1.19 g cm$^{-3}$. The blue points and error bars show the measurements.

\begin{figure}[htbp]
\begin{center}
\includegraphics[width=90mm, trim= 10 0 0 0 ]{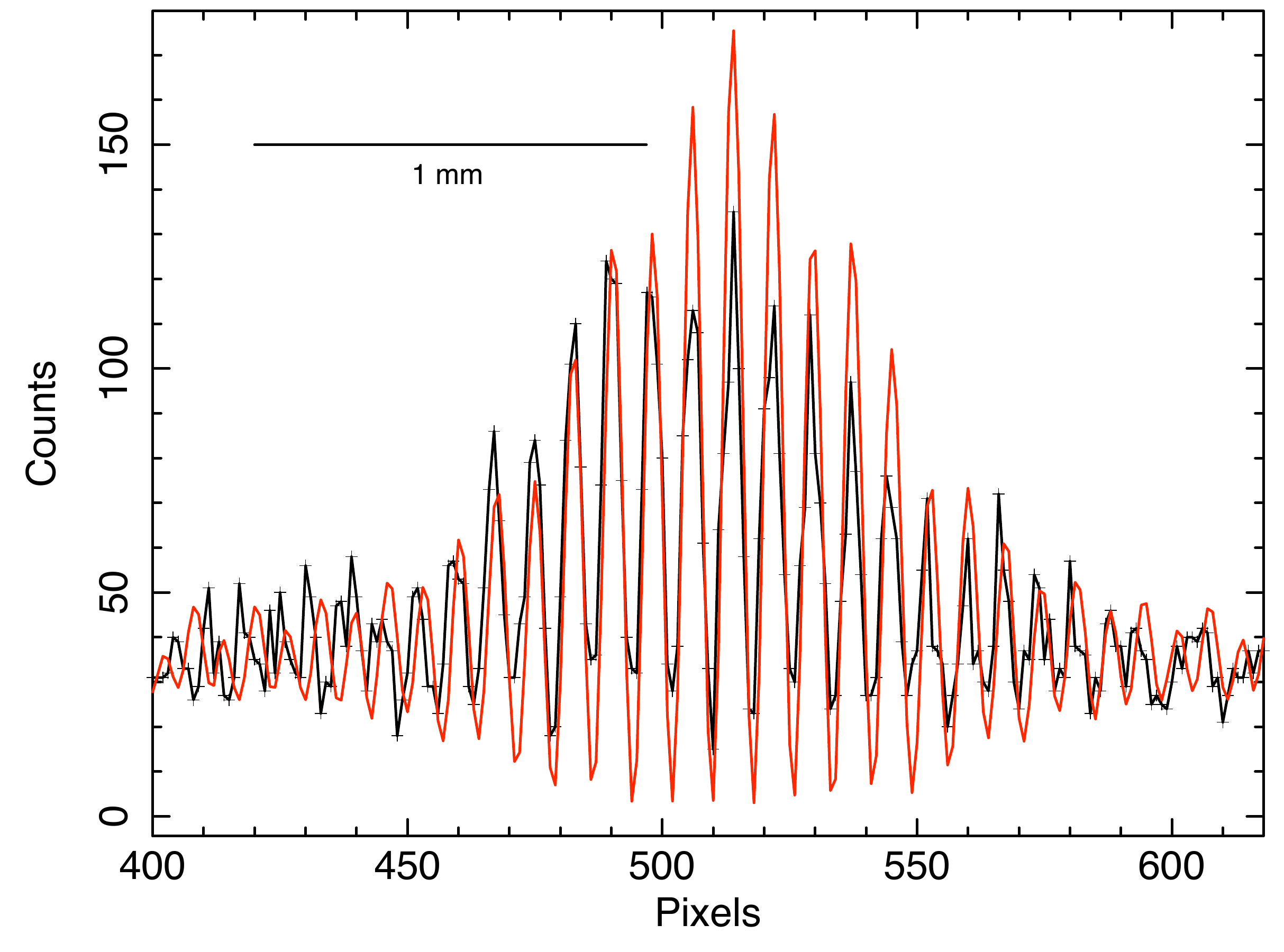}
\caption{ Projection onto the horizontal axis of fringes such as those in Fig. \ref{fig_fringes_image}, but where the detector pulse height is filtered to select only  Cu K$\alpha$ photons. The red curve is the result of a simulation. Pixels are 13 $\mu$m.}
\label{fig_fringes_diffr}
\end{center}
\end{figure}

\section{Wideband systems}
\label{sect_wideband}

With  beam combiners of limited transverse extent, as shown in Fig. \ref{fig_generic},  fringes are formed only over the limited range in $f$ over which the beams cross.  With mirrors or mirror assemblies as beam combiners, this  range is independent of wavelength, but for either refractive and diffractive combiners that is no longer the case.  Looked at another way, for a detector plane at a fixed distance  $f$, the region that has to be populated with beam combiner becomes a function of wavelength. In the diffractive case it is the region at an off-axis distance  $y=f\lambda/p$ and the extent  $\Delta y$ of the combiners places a limit $\Delta E/E \sim \Delta y /y$ on the effective energy range.

\begin{figure}[htbp]
\begin{center}
\includegraphics[width=82mm, trim= 0 0 0 0 ]{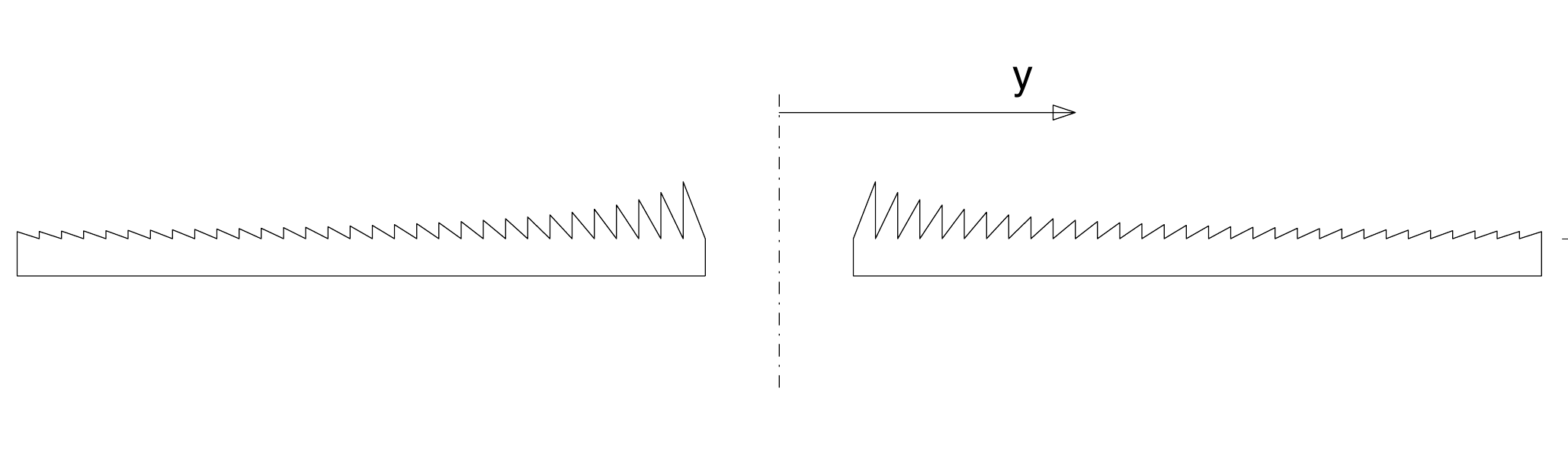}
\caption{ Profile of a pair of diffractive beam combiners designed to cover a wide energy band. The sawtooth depth at a given distance, $y$,  from the centre is made equal to \ttwopi\  for the wavelength that will be diffracted onto the detector from that region. Thus  it  varies inversely with the distance from the axis. }
\label{fig_profile}
\end{center}
\end{figure}

When extending a diffractive beam combiner in the $y$ direction so that fringes are formed for a wider range of wavelengths on a detector at the same position, the best performance will be obtained if the grating at a given $y$ is blazed for the wavelength at which that section is producing fringes. This is achieved when the profile depth is everywhere equal to the thickness of material that changes the phase bv 2$\pi$, a parameter termed  \ttwopi\   in  \citet{\skinnerpfla}.   With $\delta\propto\lambda^2$ this implies a depth that   decreases as $y^{-1}$ (see Fig \ref{fig_profile}). Consideration must be given to the diffraction into orders other than the first of radiation falling onto those parts of the surface where the profile depth is not equal to \ttwopi\ for the wavelength. If the diffracting pattern covers a range of radii differing by less than a factor of two, such problems are avoided because other diffraction orders never intersect the detector.

\section{Multiple beam interferometry}
\label{sect_multibeam}

\begin{figure}[htbp]
\begin{center}
\vspace{15mm}
\includegraphics[width=90mm, trim=200 80 200 180 ]{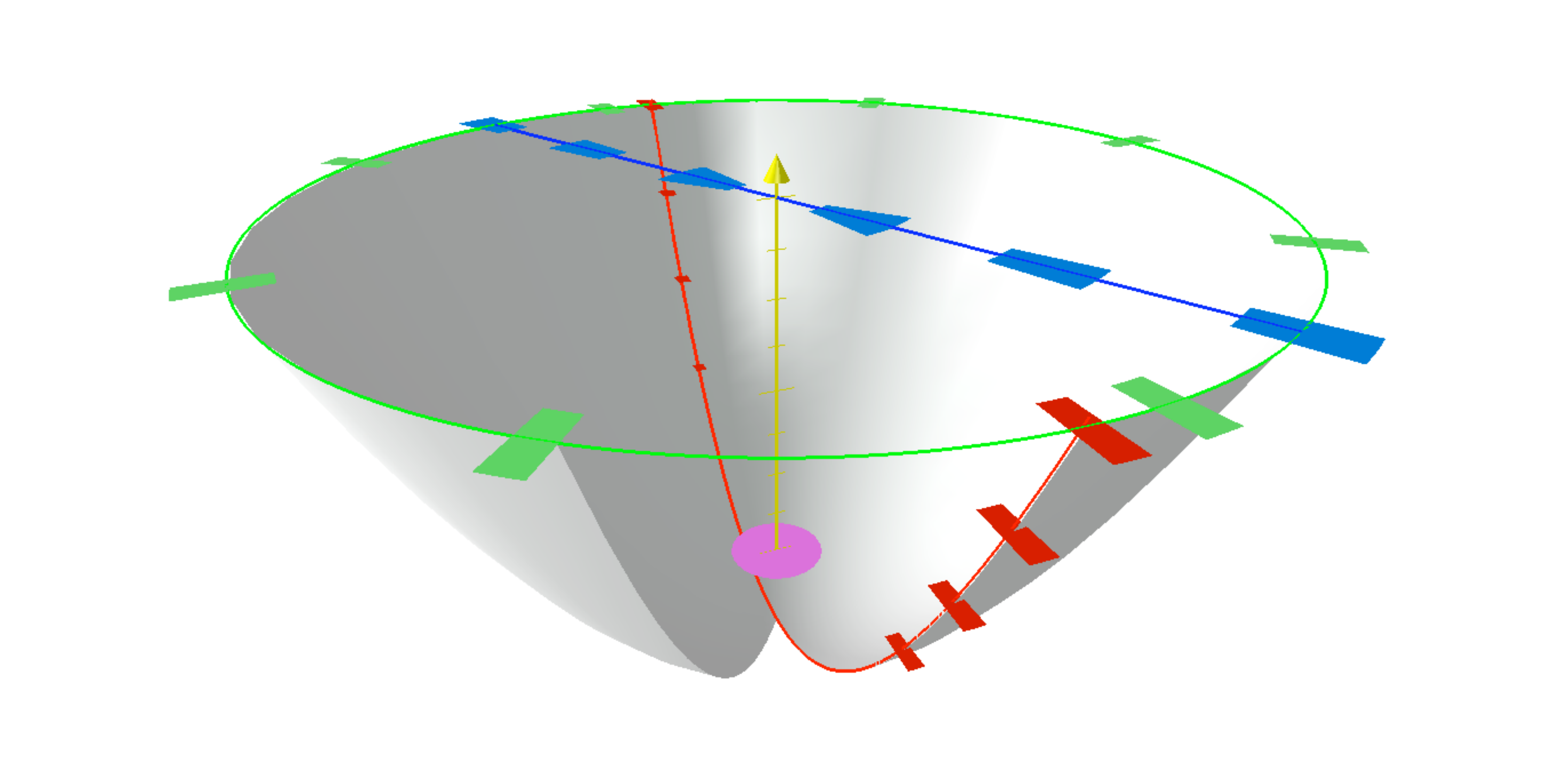}
\caption{ Possible positioning of beam combining spacecraft directing flux towards a detection plane indicated by the disk. For any position on the paraboloidal surface the total path to the detector is the same. For others, such as those indicated in blue, path compensation is needed. The special case of the positions in green is discussed in the text.}
\label{fig_paraboloid}
\end{center}
\end{figure}

We have discussed so far only combining two beams - an interferometer measuring a single vector in what in radio astronomy is referred to as the $u$-$v$ plane.
A problem arises if further beam combiners are added along the line joining the two in Fig. \ref{fig_generic} with the objective of  observing baselines with other $u$-$v$ vectors in the same direction.  Suppose an additional pair is introduced that have a pitch $p$ such that radiation of the same wavelength converges on the same detector as the first pair. Fringes will certainly be formed and if the disposition of beam combiners is symmetric the fringe pattern will also be symmetric. But because the path lengths are not the same, the  centre of the fringe pattern will not necessarily be a peak and, more importantly, its nature will be a rapid function of wavelength. For even a modest bandpass the fringe patterns from pairs of beam-combiners for which the path is the not same will not in general have the same phase and so will add incoherently, not coherently. This problem does not arise if the positions of the additional beam-combiners are limited to those on a paraboloidal surface with the detector at the focus (Fig. \ref{fig_paraboloid}). 
A special case  of potential interest occurs if the beam-combiners lie on a circle in the $y-z$ plane. In this case only the distances from a central point need be controlled with precision (Table \ref{table_tolerances}).  In the limit one could imagine a continuous ring, in which case the $u$-$v$ coverage will be a circle and the point-source response function will consequently be a $J_0^2$ Bessel function.

\section{Rotationally symmetric configurations and comparison with Fresnel lenses}
\label{sect_symmetric}

Combining the ideas of \S\ref{sect_wideband} and \S\ref{sect_multibeam} leads in the limit to a system that is rotationally symmetric with a cross-section as in Fig. \ref{fig_profile} that would be like a Phase Fresnel Lens,  but with a constant pitch. It would be an X-ray axicon that for parallel input radiation produces a line focus by launching a Bessel beam. A detector placed on the axis will see  a PSF that has almost the same form and size, independent, within wide limits, of the wavelength and position along the axis.

The similarity of  diffractive X-ray axicons  to Fresnel lenses, that have also been proposed for ultra-high angular resolution astronomy, as discussed in \S~\ref{sect_intro}, warrants a comparison of their respective capabilities and properties. 
Initially we consider a Phase Fresnel Lens (PFL) and an axicon, assuming each is perfect and ignoring absorption. The outer diameters ($d_2$) are taken to be the same, but to avoid complications associated with diffraction in orders other than +1, the   inner diameter of the axicon is taken to be $d_1=d_2/2$, while for the PFL $d_1=0$.
On these simplifying assumptions, the point source response function of the PFL and the axicon  have respectively the form of an Airy disc and (approximately) that of  $J_0^2$.
\begin{eqnarray}
I_{PFL}(\rho)=& \left(\frac{\pi d_2^2}{4 f \lambda}\right)^2\left(\frac{J_1(\rho)}{\rho}\right)^2    I_0   \label{eqn_pfl_psf}     \\
 I_{axi}(\rho) \sim & \left( \frac{2\pi^2 d_{eff}^2} {f \lambda} \right) J_0^2(\rho) I_0.
\label{eqn_axi_psf}
\end{eqnarray}
Here $I_0$ is the intensity (flux per unit area) of a parallel beam incident on the optic and $\rho=\pi q d / f\lambda$, where $q$ is the off-axis distance in the detector plane. $d_{eff}$ is the diameter of that part of the axicon that is effective for wavelength $\lambda$.
Eqn. \ref{eqn_pfl_psf}  applies exactly at the nominal energy while Eqn. \ref{eqn_axi_psf} is an approximation that applies away from the extremes of the bandpass and at $q$ that are small compared with $d_1$. The first  factor on the right hand side of each expression can be termed the  `gain', $G$, or `concentration factor'. It is the peak intensity of the image of a point source, relative to that of the incident flux. Comparing Eqns. \ref{eqn_pfl_psf}  and  \ref{eqn_axi_psf},
\begin{equation}
G_{axi}=8 \pi \left(\frac{d_{eff}}{d}\right)^2\sqrt{G_{PFL}} \: .
\end{equation}
Thus in in practical cases $G_{axi} \ll G_{PFL}$.
The  enclosed energy is respectively
\begin{eqnarray}
P_{PFL} \propto & \frac{1}{2} \left[ 1-J_0^2(\rho) - J_1^2(\rho) \right] \: \\
   P_{axi}\propto & \frac{\rho^2}{2}\left[ J_0^2(\rho) + J_1^2(\rho) \right].  \label{eqn_p_axi}
\end{eqnarray}
For increasing $\rho$ the Bessel function term in Eqn. \ref{eqn_p_axi} approaches asymptotically to $2/\pi\rho$. So, away from the central peak but within the region for which the approximations are valid,  the enclosed flux increases approximately linearly with radius.

\begin{figure}[htbp]
\begin{center}
\includegraphics[width=115mm, trim= 0 0 0 0 ]{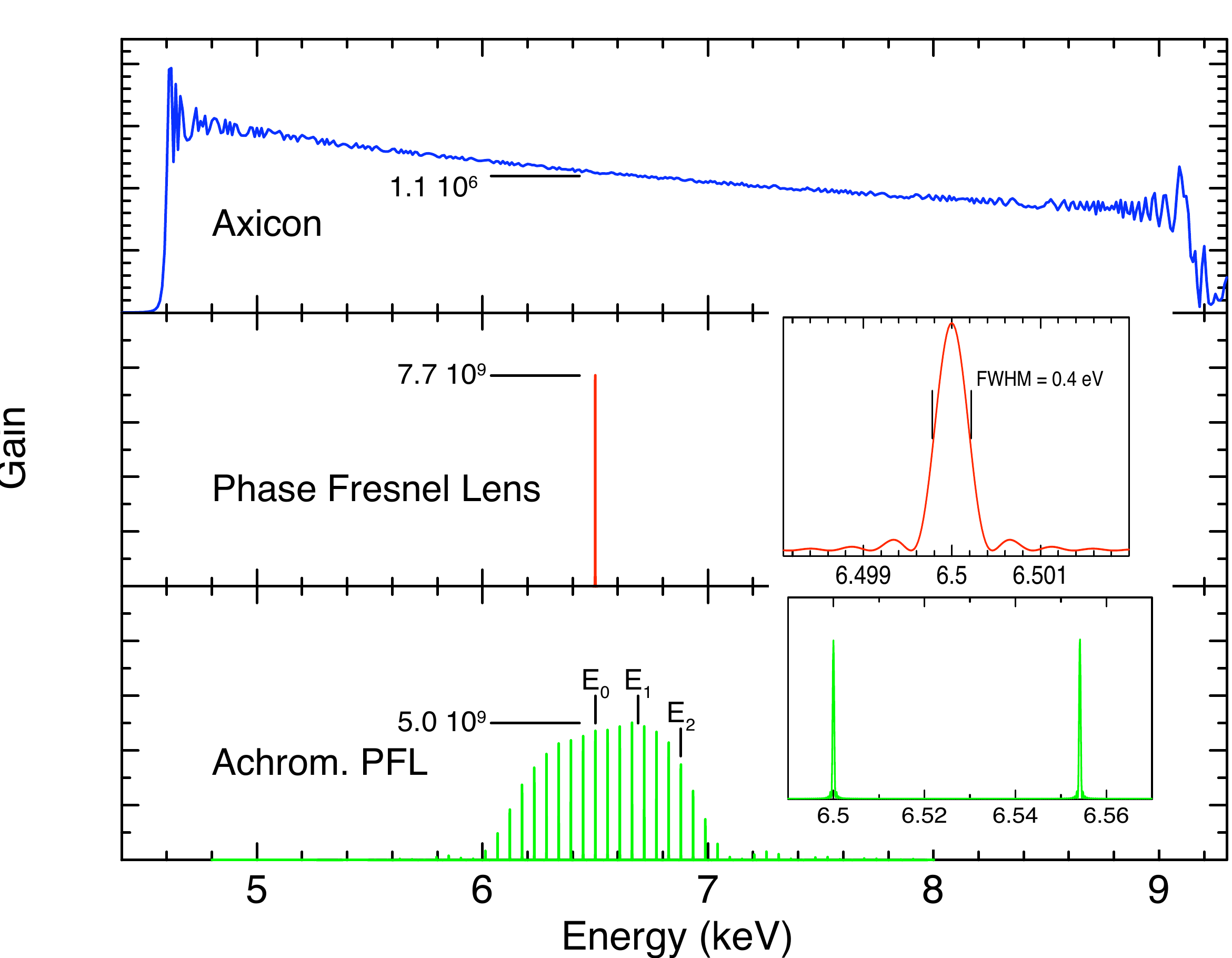}
\caption{ The `gain', $G$, or `concentration factor', defined as the peak surface brightness of the image of a point source relative to that of the incident flux, for the three optics specified in Table \ref{table_lens_egs}, as a function of photon energy. Energies $E_{0-2}$ are the energies of the example responses in Fig.  \ref{fig_encl_energy} }
\label{fig_gain_compare}
\end{center}
\end{figure}

\begin{figure}[htbp]
\begin{center}
\includegraphics[width=105mm, trim= 0 0 0 0 ]{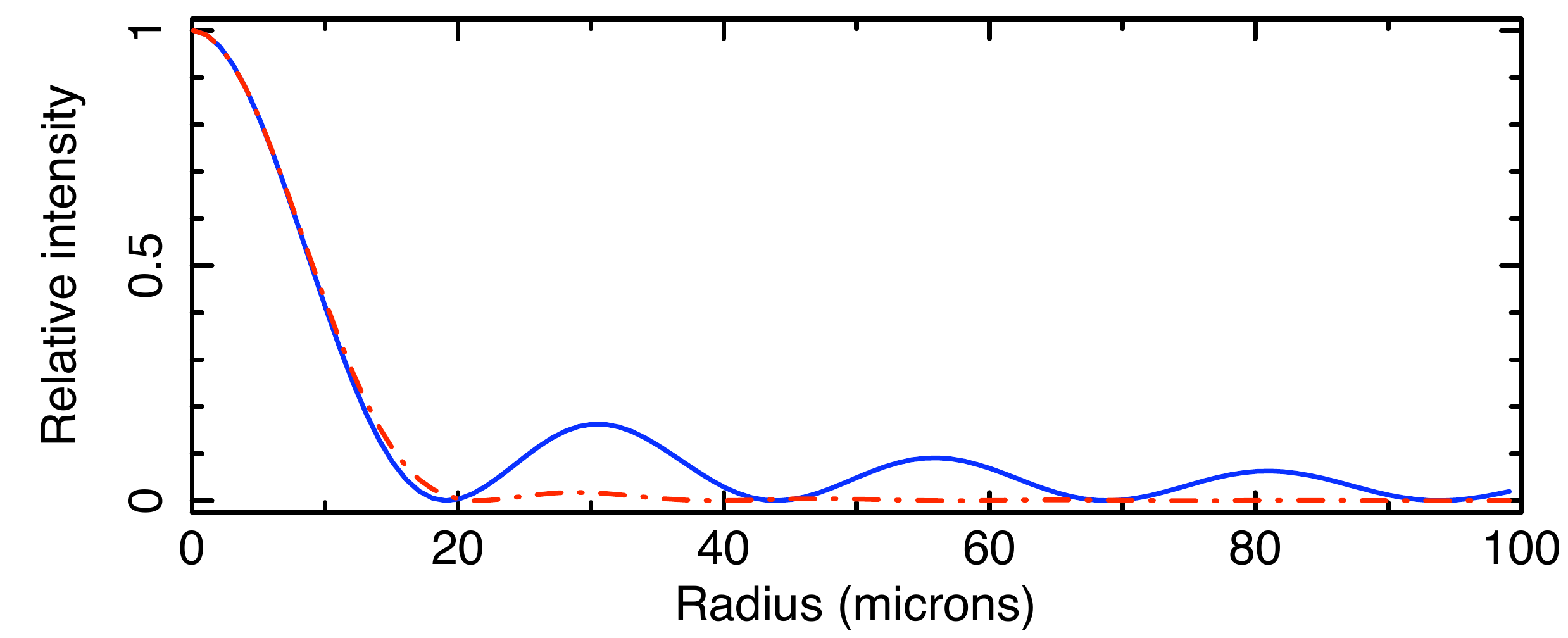}
\caption{ The radial profile of the point source response function at 6.5 keV  for the axicon specified in Table \ref{table_lens_egs} (blue) compared with that of the PFL (red). }
\label{fig_psf_shape}
\end{center}
\end{figure}

\begin{figure}[htbp]
\begin{center}
\includegraphics[width=100mm, trim= 0 0 0 0 ]{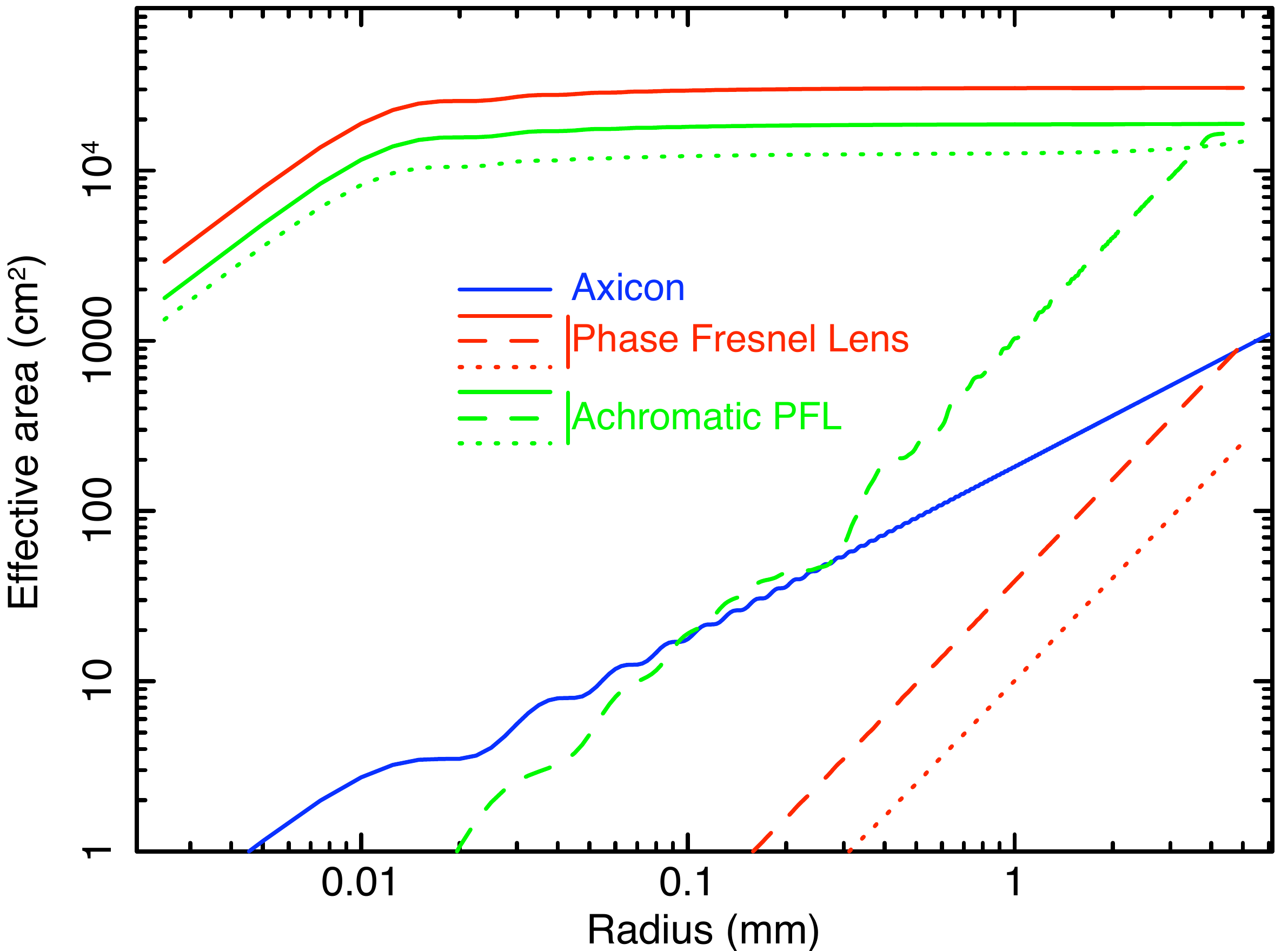}
\caption{ The enclosed energy, expressed as an equivalent effective area, as a function of radius in the detector plane for the three lenses specified in Table \ref{table_lens_egs}. Each is evaluated at  the 3 energies indicated in Fig. \ref{fig_gain_compare}: $E_0$=6.5 keV (continuous line), $E_1$=6.69 keV (dashed line)  and $E_2$=6.88 keV (dotted line). For the axicon only the $E_0$ curve is shown as the others are essentially identical.}
\label{fig_encl_energy}
\end{center}
\end{figure}

To illustrate the advantages and disadvantages of an X-ray axicon we have simulated the operation of an example design and compared it with those of a PFL and of a PFL that has been made achromatic by the addition of a stepped refractive component as suggested by various authors \citep[{e.g.}][]{\skinnerpfla}.  The assumed parameters are given in Table \ref{table_lens_egs} and results are shown in the table and in Figs. \ref{fig_gain_compare}--\ref{fig_encl_energy}.  Stepping of the Beryllium refractive lens is necessary to avoid excessive absorption and introduces the comb-like response seen in Fig. \ref{fig_gain_compare} \citep[see][]{\skinnerao}. It was taken to be such  that the profile depth never exceeds $120 t_{2\pi}$ at the nominal energy of 6.5 keV, or 2.8 mm, which is roughly the absorption length,

 \begin{table}[htdp]
\caption{The three optics  for which the simulated responses are compared in Figs. \ref{fig_gain_compare}--\ref{fig_encl_energy}. The angular resolution given is according to the Rayleigh criterion  except where  the form of the PSF precludes this, when  a HEW is specified. }
\label{table_eg_lenses}
\begin{center}
\begin{tabular}{|c|c|c|c|c|}
\hline
                     & axicon     & PFL       &  Achromatic    &  \\
                     &                   &                 &     PFL          & \\
\hline
Inner radius &    0.5        &      0        &               0       &  m \\
Outer radius&    1.0            &     1.0         &           1.0           &  m \\
Material        &  Polyimide & Polyimide &   Beryllium & \\
Pitch             &    50          &      35$^*$      &       17$^*$             & \micron \\ 
Design energy  &     9.5$^*$     &      6.5          &       6.5                   & keV    \\
Detector distance &    185    &       185                   &           185                    &         km         \\
\hline
Angular resolution&             &                           &                            &        (Rayleigh criterion)     \\
at 5 keV                     &  21           &        (31)$^ \dag$               &                           &  \muas        \\
at $E_0$=6.5 keV                 &  21            &        24                                   &          24                       &  \muas       \\
at $E_1$=6.69 keV                 &  21            &                                           &        6600 (HEW)                 &  \muas         \\
at $E_1$=6.88 keV                      &  21            &                                           &          23                  &  \muas         \\
at 8.0 keV                     &   21          &      (19)$^\dag$                &                         &  \muas         \\
\hline
\rule[-2.5mm]{0pt}{6.5mm} $\int G\: dE$      &      $4.9\:10^6$             &             $3.5\:10^6$        &      $3.4\:10^7$                         &       keV                        \\
\hline
\end{tabular}
\end{center}
$^*$ at periphery\\
$^\dag$ values for a lens of the same size designed for this energy

\label{table_lens_egs}
\end{table}%

Thus in general terms the axicon provides a much larger bandpass than even an achromatically correct Fresnel lens, though  it does so at the expense of effective area at a given energy. As the radius of the annulus that is effective at a particular energy depends inversely on energy, effective area decreases with energy. The angular resolution, according to the Rayleigh criterion is 0.765$\lambda/d_{eff}$, where $d_{eff}$ is again the diameter of the annulus diffracting radiation of wavelength $\lambda$ onto the detector. Thus measured in this way it is somewhat better than the value 1.22$\lambda/d$ for a PFL or other true imaging system. Note, however, that because of the different response shape the advantage is smaller if  FWHM  (Full Width at Half Maximum) is used as a measure. If HEW (Half Energy Width) is used then the resolution of the axicon  becomes essentially undefined because over a wide range the enclosed energy rises approximately linearly with radius.

\section{Conclusions}

We have shown that the simplicity and relatively undemanding alignment and orientation requirements of diffractive beam combiners mean that they should be seriously considered where angular resolution requirements demand long baseline interferometry at X-ray or gamma-ray energies. They provide a possible route to obtaining sub-\muas\  angular resolution with photon energies of  1--10 keV, a band that contains the important Fe lines. In view of the gravitational broadening and red-shifting of such lines when they are emitted close to a black hole, the wide band, achromatic, capabilities of instrumentation based on this  principle could be particularly valuable.

The main disadvantage is that a particular part of the diffracting surface only directs a very narrow band of energies towards the detector. Thus although  a wide  bandpass  is possible, a large surface area is required. Fortunately the diffractors are light weight (the film used in the laboratory demonstration weighs only about 200 g m$^{-2}$) and low cost.  The conceptually simplest configuration of a large, sparse, interferometric array would be with diffractors long radially, but quite narrow,  as indicated in Fig. \ref{fig_paraboloid}. Subsections with different pitch could in principle be placed side-by-side, offering a more convenient aspect ratio,  though further study would be required.

The mostly filled, rotationally symmetric, configuration -- the X-ray axicon -- is in many ways intermediate between a lens and an interferometer. The comparison with lenses presented  in Section \ref{sect_symmetric} shows that compared with lenses the central intensity (measured by $G$) is inferior by a large factor, though the bandwidth is better than that of a PFL by a comparable factor.  To collect a worthwhile flux one must, treat the system not as an imager but as an interferometer and  consider not only the energy in central peak but also that in the surrounding fringes. A useful concentration of flux does, however,  take place. In the example considered in \S\ref{sect_symmetric},  at the central energy a 1~cm$^2$ region of detector receives flux equivalent to $\sim$1000~cm$^2$ of the incoming radiation and the central peak is more than a million times brighter than the unfocussed radiation.

Further work is needed to identify the most promising regions of the design parameter space and to study variations such as radial dependencies of pitch that are intermediate between constant (axicon) and linear (PFL) or combining refractive correction.     

\begin{acknowledgements}
   
	The authors are is grateful to K. Gendreau, Z. Arzoumanian and the team responsible for the development of the GSFC  600 m interferometry test bed and to 3M  Optical Service Division and Microsharp Corporation Ltd., U.K. for providing film samples.  
	
 \end{acknowledgements}


\bibliography{beam_combiners_11.bib}   
\bibliographystyle{aabib99}                

\end{document}